\documentstyle[12pt,epsf]{article}
 \catcode`@=11
\long\def\@caption#1[#2]#3{\par\addcontentsline{\csname
  ext@#1\endcsname}{#1}{\protect\numberline{\csname
  the#1\endcsname}{\ignorespaces #2}}\begingroup
    \small
    \@parboxrestore
    \@makecaption{\csname fnum@#1\endcsname}{\ignorespaces #3}\par
  \endgroup}
\catcode`@=12
\newcommand{\bi}{\bibitem}
\begin{document}
\setlength{\baselineskip}{0.27in}

\newcommand{\beq}{\begin{equation}}
\newcommand{\eeq}{\end{equation}}
\newcommand{\beqa}{\begin{eqnarray}}
\newcommand{\eeqa}{\end{eqnarray}}
\newcommand{\lsim}{\begin{array}{c}\,\sim\vspace{-21pt}\\<
\end{array}}
\newcommand{\gsim}{\begin{array}{c}\sim\vspace{-21pt}\\>
\end{array}}

\begin{titlepage}
{\hbox to\hsize{Dec. 1995\hfill } }
{\hbox to\hsize{UM-TH-95-15\hfill}}
{\hbox to\hsize{UCSD/PTH 95-22\hfill}}
\begin{center}
\vglue .06in
{\Large \bf The Extraction of $V_{ub}$ from Inclusive B Decays  and  the  Resummation  of End Point Logs}
\\[.5in]
\begin{tabular}{c}
{\bf R. Akhoury}\\[.05in]
{\it The Randall Laboratory of Physics}\\
{\it University of Michigan}\\
{\it Ann Arbor MI 48109 }\\[.15in]
\end{tabular}
\vskip 0.1cm
\begin{tabular}{c}
{\bf I.Z. Rothstein}\\[.05in]
{\it Dept.  of Physics}\\
{\it UCSD}\\
{\it La Jolla Ca. 92093 }\\[.15in]
\end{tabular}
\vskip 0.25cm

{\bf Abstract}\\[-0.05in]

\begin{quote}
In this paper we discuss the theoretical difficulties in extracting $V_{ub}$
using the data from inclusive B decays. Specifically, we address the issue
of the end point singularities. We  perform the resummation
of both the leading and next to leading end point logs and
 include the leading corrections to the hard scattering amplitude. 
We find that the resummation is a $20\%-50\%$ effect in the end point
region where the resummation is valid. Furthermore, the resummed
sub-leading logs dominate the resummed  double logs.
 The consequences of this
result for a model independent extraction of the mixing angle
$V_{ub}$ are explored.

\end{quote}
\end{center}
\end{titlepage}
\newpage
\section{Introduction}
 
Measurements in the bottom quark sector have reached the point that
our knowledge of many observables is now bounded by the theoretical
uncertainties \cite{rb}.
Fortunately, theoretical advances in calculating both exclusive as well
as inclusive rates now allow the extraction of the CKM parameters without 
recourse to the models which have soiled the extraction processes to data.
 The present values of $V_{ub}$ 
have a model dependence
which introduce an uncertainty of a factor of 2\cite{rb}, which is several
times larger than the experimental uncertainties.
With QCD based calculations, we can now hope to extract both $V_{bc}$
and $V_{ub}$ with errors on the order of tens of percents. In this work,
we concentrate on the extraction of $V_{ub}$ from the measurement of the electron spectrum in semi-leptonic 
inclusive B meson decays.

The extraction of $V_{ub}$ from inclusive semi-leptonic B decays
is hindered by the fact that the background from charmed decays
is overwhelming for most of the range of the lepton energy.
Thus, we are forced to make a cut on the lepton energy, vetoing all events,
or some large fraction thereof, with lepton energy less than
the $b\rightarrow c$ end point energy. Given the proximity of the two relevant
end points, this obviously hinders the statistics. However, even with a
large data sample, the accuracy of the extraction will be limited by the
errors induced from the approximations used in calculating the 
theoretical prediction in the end point region. This region of the
Dalitz plot is especially nettlesome for theory, because the perturbative, as
 well as the non-perturbative corrections become large when the lepton
energy is near its endpoint value.

It has been shown that it is possible to calculate the decay spectrum
of inclusive heavy meson decay in a systematic expansion in
$\epsilon=\frac{\Lambda_{QCD}}{m_b}$ and $\alpha_s$ using an operator
product expansion within the confines of heavy quark effective field
theory\cite{cgg}. It is possible to Euclideanize the calculation of the rate
for most of the region of the Dalitz plot with only minimal assumptions
about local duality. However, in the end point region, the expansion
in $\epsilon$, as well  as the expansion in $\alpha_s$, begin to breakdown
(The endpoint region poses problems for local duality as well. We shall
discuss this in more detail later).

The aim of this paper is to determine the size of the errors induced
from the theoretical uncertainties in extraction of $V_{ub}$.
A large piece of this work consists of implementing 
the resummation of the
leading and sub-leading endpoint logs which cause the breakdown of the
expansion in $\alpha_s$, as first discussed on general grounds 
in \cite{sk}, and the inclusion
of the $\alpha_s$ corrections to the hard scattering amplitude. 
However, to determine the consistency of our
calculation, we must also address the issue of the non-perturbative
corrections. These issues have been previously looked at in
refs \cite{fjmw} and \cite{bm}. In \cite{fjmw} the need for resummation
was addressed on general grounds. However, the calculational methods
used here are not compatible with the arguments given in
\cite{fjmw}, and thus we must recapitulate these arguments within
the confines of our methods.

In the second section of this paper, we discuss the question of the need
 to resum the perturbative as well as non-perturbative series.
The next three sections are dedicated to the resummation of the leading
and next to leading infrared logs and the inclusion of the one
loop corrections to the hard scattering amplitude 
(read one loop matching). In the fifth section we
give our numerical results while the last section draws conclusions 
regarding what errors we can expect in the extraction process.   

\section{Is Resummation Necessary?}

As mentioned above, the theoretical calculation of the lepton spectrum
in inclusive decays breaks down near the endpoint. Both the non-perturbative
as well as perturbative corrections become large in this region.
Here we investigate the need to perform resummations in either or both 
of these expansions. The one loop decay spectrum including the leading
non-perturbative corrections is given by \cite{mw}

\begin{eqnarray}
\label{oneloop}
\frac{1}{\Gamma_0}\frac{d\Gamma}{dx}&=&
\theta(1-x)
\left[ x^2(3-2x)(1-\frac{2\alpha_s}{3\pi})I(x)
+2(3-x)x^2
E_b-\frac{2}{3}x^2(9+2x)K_b-\frac{2}{3}x^2(15+2x)G_b 
 \right] \nonumber \\
&+&\left[  E_b-\frac{2}{3}K_b+\frac{8}{3}G_b \right] \delta(1-x)+\frac{1}{3}K_b
\delta^{\prime}(1-x).
\end{eqnarray}
Where
\beq
I(x)=\log^2 (1-x)+\frac{31}{6}\log (1-x)+\pi^2+\frac {5}{4}~~{\rm{and}}
~~x=\frac{2 E_e}{m_b},
\eeq
\beq
\Gamma_0=\mid V_{ub}\mid^2\frac{G_F^2m_b^5}{96 \pi^3}.
\eeq
$E_b,~G_b$ and $K_b$ are hadronic matrix elements of order $\epsilon^2$
and are given by
\begin{eqnarray}
E_b&=&G_b+K_b, \nonumber \\
K_b&=&\langle B(v)\mid \bar{b}_v\frac{D^2}{2m_b^2}\mid B(v)\rangle,\nonumber \\
G_B&=&\langle B(v)\mid \bar{b}_vg\frac{\sigma_{\alpha \beta}G^{\alpha \beta}}
{4m_b^2}b_v\mid B(v)\rangle,
\end{eqnarray}
$b_v$ is the velocity dependent bottom quark field as defined in
heavy quark effective field theory.
From the above expressions we see that the breakdown of the expansions, in
$\alpha_s$ and $\epsilon=\frac{\Lambda_{QCD}}{m_b}$, manifest themselves in the
large logs and the derivative of delta functions, respectively.

\subsection{The non-perturbative expansion}
As one would expect for heavy meson decay, the leading order term in
$\epsilon$ reproduces the parton model result. All corrections due to the fact that
the b quark is in a bound state are down by $\epsilon^2$
\cite{cgg}. 
However, near the end point of the electron spectrum we begin to probe the non-perturbative
physics.
 The general form of the expansion in $\epsilon=\frac{\lambda}{m_b}$, to
leading order in $\alpha_s$, is given as follows
\begin{eqnarray}
\label{OPE}
{1\over {\Gamma_0}}{d\Gamma\over{dx}}& = &\theta(1-x)\left(\epsilon^0+\epsilon^2+\cdot \cdot \cdot \right)+\delta(1-x)\left(0 \epsilon +
\epsilon^2+\epsilon^3+\cdot \cdot \cdot\right) \\
&+&\cdot \cdot \cdot+\delta^{(n)}(1-x)\left(\epsilon^{n+1}+\epsilon^{n+2}
+\cdot \cdot \cdot \right)+\cdot \cdot \cdot
\end{eqnarray}

The end point singularities are there because the true end point
is determined by the mesonic mass and  not the partonic mass, as enforced
by the theta function in the leading order term. 
 The difference between these end points
will be on the order of a few hundred MeV.
To make sense of this expansion we must smear the decay amplitude with some smooth
function of $x$. Normally, this would not pose a problem, however,  given that the distance between the $b \rightarrow c$ and $b\rightarrow u$ end points is approximately $330~MeV$,
we are forced to integrate over a weighting function which has support 
in a relatively small region. On the other hand, if the  weighting function is too narrow, then the expansion in  $\epsilon$ 
will not be well behaved.

Thus we must find a smearing function that minimizes the errors due to 
${\Lambda_{QCD}\over{m_b}}$ corrections which does
not  overlap with the energy region where we expect many $b\rightarrow c$ events.
The question then becomes how many $b\rightarrow c$ transitions can we allow without
introducing large errors due to our ignorance of the $b\rightarrow c$ end point spectrum (the theory breaks down in the $b\rightarrow c$ end point region as well though there are important differences between this case
and the $b\rightarrow u$ transitions)?

The issue of smearing  was addressed by Falk et. al.\cite{fjmw} who  used Gaussian smearing functions to gain quantitative insight into the need for smearing.
They found that without any resummation, the smearing function should
have a width which is greater than $\epsilon$, but that after resumming the
leading singularities, we need smear only over a region of
width $\epsilon$. 
Here we will smear by taking moments of the electron energy spectrum
(we work with the moments of the spectrum because it greatly facilitates the
resummation of the perturbative corrections). Thus, we must address the question of what range of values of $N$ will lead to a sensible expansion  which is also
not overly contaminated by $b\rightarrow c$ transitions?
This will obviously depend on the ratio of $V_{ub}$ to $V_{bc}$.
To get a handle on the numerics, let us for the moment assume that we
wish that the number of $b\rightarrow u$ transitions be at least equal
to the number of $b\rightarrow c$ transitions in our sample.
In figure 1.,  we plot $\frac{V_{ub}^2}{V_{bc}^2}(N)$, which is  the ratio of mixing angles for which the $N$th moments of the leading
order rates
for $b\rightarrow c$ and $b\rightarrow u$ transitions will be equal, and
is given by
\begin{equation}
\frac{V_{ub}^2}{V_{bc}^2}(N)=
\frac{ \displaystyle 
\int^{x_m}_0x^2{(x_m-x)^2 \over{(1-x)^3}}[6-3x_m+(x_m-6)x+2x^2]x^Ndx}
{\displaystyle \int^1_0x^2(3-2x)x^N}.
\end{equation}
$x_m$ is $\frac{2E_{max}}{m_b}$ for the $B\rightarrow D$ transitions
and takes the value $x_m\approx.9$. 
Given the bounds \cite{pdt} 
\begin{equation}
.002<\frac{\mid V_{ub} \mid^2}{\mid V_{bc} \mid^2}<.024,
\end{equation}
we see that an understanding of the spectrum for moments around $N\simeq20$
is necessitated if we wish to keep the $b\rightarrow c$ contamination
under control (Of course we do not suggest that these moments can be measured
given the finite resolution of the experiment. We will discuss this
situation later in the paper).

\begin{figure}
\centering
\epsfysize=2.5in   
\hspace*{0in}
\epsffile{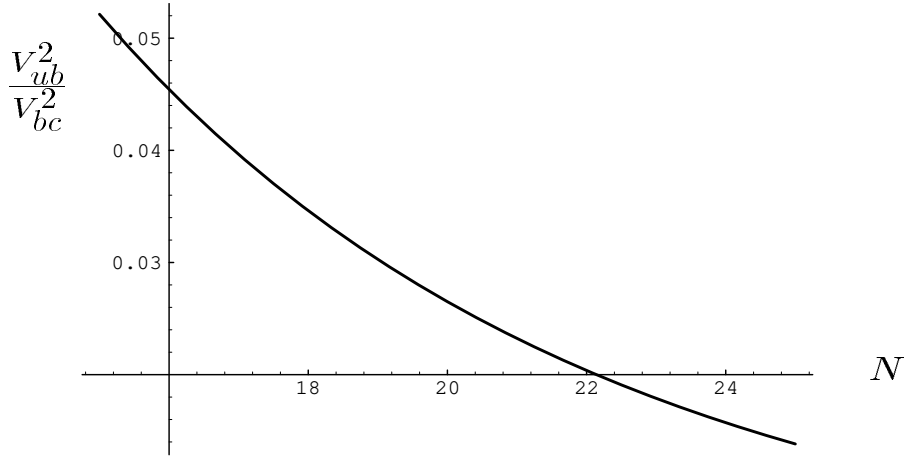}  
\caption{The ratio of $\frac{V_{ub}^2}{V_{bc}^2}$ for which the Nth moments
of the leading order spectra are equal.}
\label{vub(n)}  
\end{figure}

We now consider the issue of determining the maximum value
of $N$ for which the expansion makes sense. Let us first consider the expansion in $\epsilon$. The moments of the leading singularities of eq.(\ref{OPE}) will behave as
\beq
M_N \approx C_n \frac{N!\epsilon ^{n+1}}{(N-n)!} .
\eeq
As a possible criterion on the size of $N$, we may impose that there be no
growth with n. That is
\beq
\frac{N!\epsilon^{n+1}\left( ln\epsilon+\Psi (N-n+1)\right)}{(N-n)!}<0.
\eeq
$\epsilon$ represents the value of some matrix element in the
heavy quark effective theory.
It is assumed that the value of $\epsilon$ should be 
on the order of a few hundred $MeV/m_b$, but in theory it
could vary by a factor of order one from term to term.
To get a handle on the sizes of $\epsilon$, we may consider the leading 
$\epsilon$, 
which is given by
\beq
\epsilon_1=\langle B \mid \bar{b}_v\frac{(iD)^2}{2}b_v \mid B\rangle \frac{1}{m_b^2}.
\eeq
Quark model calculations suggest that $\epsilon_1^2$ is on the order
of .01\cite{neub}. Thus, naively, it seems that to keep the expansion in 
$\epsilon$ under control, we must keep $N\leq 10$.
This estimate
is perhaps too crude for our purposes given that we know nothing
of the growth of the coefficients $C_n$ nor of the range
of possible values of $\epsilon_N$, it does suggest that some sort
of resummation may  be necessary. 

  Neubert\cite{neub} pointed out that it is possible to resum the leading singularities, much
as in the case of deep inelastic scattering, into a non-perturbative shape function
\beq
\tilde{f}(k_+)=\langle B \mid \delta(k_+-iD_+)\mid  B \rangle.
\eeq
This function gives the probability to find the $b$ quark within the hadron with residual light cone
momentum $k_+$. Thus, this function is roughly determined by the kinetic
energy of the b quark inside the meson.
This structure function will be centered around zero and have some
characteristic width $\delta$. $\delta$ will determine the maximum
size of $N$ for which the expansion without resummation makes sense.
To get a well behaved expansion we choose $N$ such that $x^N$ gives order one
 support to the structure
function throughout its width. The value of $\delta$ is unknown 
at this time, and various authors
have given different estimates for its value. We can assume that this
width should be on the order of $(m_B-m_b)/2$ which is around 
300 $MeV$. We shall choose, what we believe to be the conservative 
value of 500 $MeV$ for $\delta$. Since the structure function is the
sum of derivatives of delta functions, we conclude that we should smear
over the width of the function if we do not wish to incur large errors.
Let us assume, for the sake of numerics, that $x^N$ should not fall
below the value .1, within 500 $MeV$ of the end point.
Then we find that $N$ must be $\leq20$.  
Thus, we expect the non-perturbative effects could be quite large for the
range of $N$ that we consider here.
Of course when $N$ becomes very large, $N>100$, it is necessary to go
beyond leading twist since the soft gluon exchange in the t channel
begins to dominate, not to mention the failure of the OPE due to
its asymptotic nature \cite{shif}.

We see that  for our purposes we should include the non-perturbative
structure function in our calculation. 
The fact that the knowledge of this non-perturbative 
function is needed to extract $V_{ub}$
should not bother us too much however, given that it is universal.
That is to say we can remove it from our final result by taking the
appropriate ratio \cite{bigmat}. 
Or it can be measured on the lattice, much in the same
way that the moments of the proton structure functions are now
being measured. Using the ACCMM model \cite{accm} Blok and Mannel \cite{bm}
concluded
that a resummation of the non-perturbative corrections is unnecessary.
If the width of the structure function is smaller than the conservative
number chosen here, then this could very well be true.
This would be a welcomed simplification of  the extraction process,
since we would no longer need to rely on the extraction of 
non-perturbative parameters from other processes to
measure the mixing angle $V_{ub}$.

\subsection{The perturbative expansion}  
Let us now address the issue of the perturbative corrections.
 The corrections in $\alpha_s$ grow large near the electron energy
end point, and, precisely at the end point,  there are logarithmic infrared
divergences. These divergences are due to the fact that near the end point
gluon radiation is inhibited, and as a result, the usual cancelation
of the infrared divergences between real and virtual gluon emission is
nullified. Of course, the rate is not divergent, and we expect that
a resummation procedure will have the effect of reducing the rate for the 
exclusive
process.

  Near the end point large logs form a series  of the form
\begin{eqnarray*}
\label{psum}
{d\Gamma\over{dx}}&=& C_{11}\alpha Log^2[1-x]+C_{12}\alpha Log[1-x]+C_{13}\alpha \\
                  &+& C_{21}\alpha^2 Log^4[1-x]+C_{22}\alpha^2 Log^3[1-x]+\cdots\\
                  &+& C_{31}\alpha^3Log[1-x]^6+C_{32}\alpha^3 Log^5[1-x]+\cdots\\ 
                  &+&~~~~~~~~~~~.~~~~~~~~~~~+~~~~~~~~~.~~~~~~~~~~~~+\cdots\\
                  &+&~~~~~~~~~~~.~~~~~~~~~~~+~~~~~~~~~.~~~~~~~~~~~~+\cdots\\
\end{eqnarray*}
Which in terms of a moment expansion gives (for large $N$)\footnote{This form
holds for $b\rightarrow s\gamma$, for the semi-leptonic decay we will consider
the moments of the derivative of the rate.}

\begin{eqnarray}
\label{momsum}
\int^1_0 dx x^N{d\Gamma\over{dx}}=&& \frac{1}{N}\left(\tilde{C}_{11}
\alpha Log^2N+\tilde{C}_{12}\alpha LogN
+\tilde{C}_{13}\alpha \right. \nonumber \\
                  &+& \tilde{C}_{21}\alpha^2 Log^4N+\tilde{C}_{22}
\alpha^2 Log^3N+\cdots
\nonumber \\
                  &+&\left.
                  \tilde{C}_{31}\alpha^3LogN^6+\tilde{C}_{32}\alpha^3 
Log^5N+\cdots\right).
\nonumber \\ 
\end{eqnarray}

Given this expansion, we may ask what errors we expect to incur by
truncating the expansion at order $\alpha_s$? For $N$ near 20, we see that
\beq
\label{constraint}
{\alpha_s\over{\pi}}Log^2N\simeq .6,
\eeq
so we might expect that truncating at leading order would not be such
a good idea. We must also note that in (\ref{oneloop}) the
sub-leading log actually dominates the leading double log due to
the large coefficient $\frac{31}{6}$. The resummation of the double logs
is simple and leads the the exponentiation of the double logs.
Figure 2 shows the difference
between the one loop result and the result with only the double
logs resummed. We see that the difference is very small, on the order
of five percent. Thus, one might come to the conclusion that no resummation
of the perturbative series is necessary. 
However,  given that the coefficients of  the single logs as well
as the $\pi^2$, which are just as large as the double logs for the range of
$N$ we are considering here, are unknown at higher orders, we can only 
determine the errors induced  by a truncation of the series
after we have performed the resummation.

\begin{figure}
\centering
\epsfysize=2.4in   
\hspace*{0in}
\epsffile{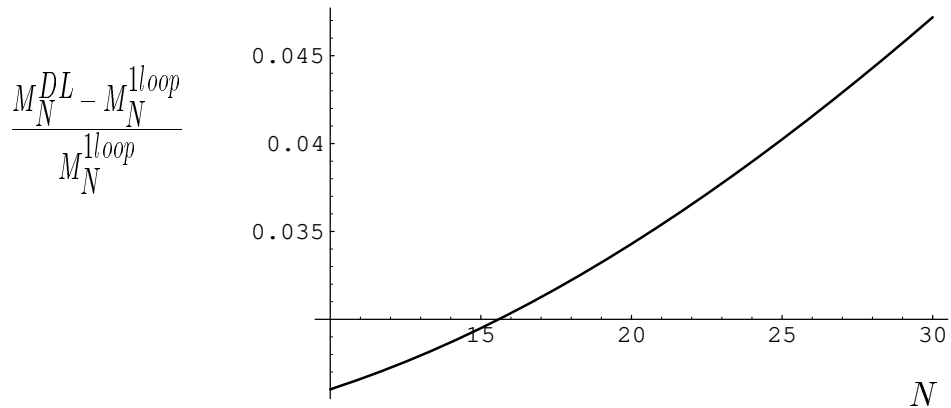}  
\caption{The difference between the moments given by the leading
$\alpha_s$ correction and the moments of the rate with only the double
logs resummed.}  
\end{figure}

Resumming the leading double logs 
in itself does not increase the range in $N$
over which perturbation theory is valid. 
Even after this resummation is performed the criteria for a convergent
expansion is still $\frac{\alpha_s}{\pi} Log^2N<1$
unless we know that the subleading logs exponentiate as well.
However, one can show on very general grounds \cite{sterman}
that all the end point
logs exponentiate as a consequence of the fact that these logs
are really just UV logs in the effective field theory\cite{ft,kr}.
Thus it is always possible to write down a differential equation for
the rate based on its factorization scale independence. 
As such, the general form of the decay amplitude will be given by
\beq
\label{exp}
Log\left[ \int^1_0 \frac{d\Gamma}{dx}x^N\right]=C(\alpha_s)+\sum_{n=1}
^{\infty }\alpha^n_s\sum^{2n}_{m=1}G_{nm}Log^mN .
\eeq
Once we have this information, the question of the region of convergence
becomes, do the lower order terms in question contribute numbers
of order 1 in the exponent? We may continue to increase $N$ until
we find that the subleading terms in the exponent contribute on the 
order 1. 
Thus in general resumming the leading logs does indeed allow us to
take $N$ into the range where $\frac{\alpha_s}{\pi}Log^2N\simeq 1$.
Here we will go further, as was done in \cite{sk}, and sum the next to leading
logs as well, allowing $\frac{\alpha_s}{\pi}LogN\leq 1$.
This will allow us to determine the convergence of the expansion.
Furthermore,
we extend the analysis of \cite{sk} to include the one loop matching corrections thus completing the calculation at order $\alpha_s$.

We wish to note that Blok and Mannel \cite{bm} analyzed the effects
of the large logs to the end point spectrum and concluded that no
resummations were necessary. These authors propose to take the lower
bound on the moment integral to be the charmed quark endpoint $x_c$.
Doing this allows one to stay away from larger values of $N$ (the authors
choose $N<10$). Cutting off the integral introduces errors that have the doubly
logarithmic $x_c$ dependence $ln^2(1-x_c)$. To reduce these errors, it is 
necessary to go to higher values of $N$. These authors claim that for
$N<10$, the errors induced by cutting off the integral are small, on the
order of a few percent.
However, we believe that these authors have underestimated their errors
because they normalized their errors by the total width and not
the moments themselves. Furthermore, and perhaps most importantly, 
the authors did not consider the
possibility that the sub-leading logs could dominate the leading
logs in the resummation, which as we shall see, is indeed the case.

Finally, it should be pointed out that aside from being bounded
by the size of the logs, $N$  is bounded on purely logical grounds.
The whole perturbative QCD framework loses meaning when 
the time scale for gluon emission becomes on the order of the
hadronization time scale. This restriction bounds  the
minimum virtuality of the gluon, which we expect to be on the 
order of $\frac{m_b}{N}$(we will show this to be true when we perform the
resummation). Thus, performing
resummations can only take one so far no matter how powerful one is.
However, for top quark decays it is possible to get extremely close to
the end point due to the large top quark mass. In this case it is clear
that the resummation of the next to leading logs will  become
essential. Thus, the extraction of $V_{td}$ from inclusive top quark decays
will have much smaller theoretical errors than in the b decay case.
We shall discuss the issue of the breakdown of perturbation theory
in greater detail after we perform the resummation.

\section{Factorization}

The large logs appearing in the perturbative expansion arise from the
fact that at the edge of phase space gluon emission is suppressed.
The problem of summing these large corrections has been treated previously
for various applications, such as deep inelastic scattering and
Drell-Yan processes\cite{sterman,ct}, just to mention a few.
The case of inclusive heavy quark
decay has been treated previously in \cite{sk}.
An important ingredient of the
resummation procedure is the proof of factorization.
 As applied to the present processes,
this procedure separates the particular differential rate under consideration 
into sub-processes with disparate scales. 

 In the case of inclusive semi-leptonic heavy quark decays, the relevant scales are $m_b$ and  
$m_b(1-x)$, with $x= {2E_e\over{m_b}}$ in the rest frame of the b-quark.
To understand how to best factorize the differential rate in the limit
$x \rightarrow 1$, we need to know the momentum configurations which 
give leading contributions in that limit. With this in mind, 
let us consider the inclusive decay of the b-quark into an
electron and neutrino of momenta $p_e$ and $p_\nu$ respectively,
and a hadronic jet of momenta $p_h$.
First we note that
 the kinematic analysis is simplified with the following choice of variables
in the rest frame of the b quark\cite{svb} 

\beq
x={2E_e\over{m_b}}~~~y_0={2(E_e+E_\nu)\over{m_b}}~~~y={(p_e+p_\nu)^2\over{m_b^2}}.
\eeq
The kinematic ranges  for these variables are
\beq
0 \leq x \leq 1;~~ 0 \leq y \leq x;~~ ({y/x}+x) \leq y_{0} \leq (y+1).
\eeq
Furthermore,  define the variable
\beq
\eta=\left({1-y\over{2-y_0}}\right)~~\rm{where}~~x\leq\eta\leq 1.
\eeq
This variable plays an analogous role to the
 Bjorken scaling variable in deep inelastic scattering phenomena. 
The invariant mass of the final state hadronic jet, and its energy are 
given by

\beq
p_h^2=m_b^2(1-\eta)(2-y_0),~~~ p_{h}^{0}=\frac{m_b}{2}(2-y_0).
\eeq
We should note that in determining the boundary values of the various variables we
refer to the b-quark mass and not to that of the meson. This is justified
within the perturbative framework we are working in at the moment.
However, once we include the effects of the non-perturbative structure
function, the phase space limits will take on their physical values.

Let us now investigate the dominant momentum configurations near 
$x \rightarrow 1$. First, we observe that the invariant mass of the hadronic jet
$+$ neutrino system is given by
$(p_b-p_e-p_\nu)^2={m_{b}^2}(1-x)$ which vanishes at the end point. The phase 
space configuration where the neutrino is soft is suppressed and hence,
when the value of $x$ approaches one,  the electron and the hadronic jet-neutrino
system move back to back in the rest frame of the b-quark. Furthermore,
the invariant mass of the hadronic jet  
vanishes independently of the neutrino energy. This is readily verified using the phase space
boundaries. The energy of the jet is large except near the point
$x \rightarrow y \rightarrow 1$. In this region of the Dalitz plot factorization breaks down, and the techniques used here fail. However,
 this problematic region is irrelevant as a consequence of the fact that 
 the rate to produce soft massless fermions are suppressed at the tree level.
Thus, the following picture emerges at $x \sim 1$. The b-quark decays
into an electron moving back to back with the  neutrino and 
a light-like hadronic jet.  We choose the electron 
to be moving in the $+$ (light cone) direction, and  the jet moves in the $-$ (light cone) direction
in the rest frame of the b-quark. 
 The constituents of the jet may interact via
soft gluon radiation with each other and with the b-quark, but hard
gluon exchange is disallowed. 

\begin{figure}
\label{reduced}
\centering
\epsfysize=3in   
\hspace*{0in}
\epsffile{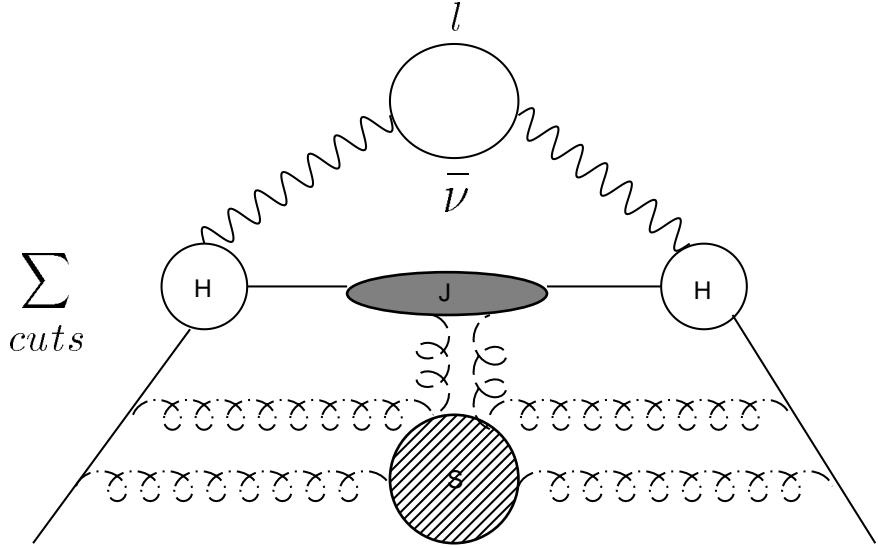}  
\caption{Reduced diagram for B decays.}
\label{rbos}  
\end{figure}

This simple picture is related by the Coleman-Norton theorem \cite{cn,s}
 to the type
of Feynman diagrams that are infrared sensitive. According to this theorem,
 if we  construct a ``reduced'' diagram by contracting all off shell lines to a point, then at the
infrared singular point, such a diagram describes a physically
realizable process. Thus at $x \sim 1$ the type of diagrams that
give large logs are precisely those described above and shown in figure \ref
{reduced}.
 
In the figure, {\it{S}} denotes a soft blob which interacts with the jet and the
b-quark via soft lines. {\it{J}} denotes the hadronic jet and {\it{H}} the hard scattering
amplitude.  The typical momenta flowing through the hard  sub-process
are $O(m_b)$. {\it{H}} does not contain any large end point logs and has a
well defined perturbative expansion in $\alpha_s(m_b)$. All the lines which constitute
{\it{H}} are off-shell and have been shrunken to a point. The soft function
{\it{S}} contains 
typical soft momentum $k$ , with 
$k^{+} \sim k^{-} \sim k_{\perp} \sim O(m_{b}(1-x))$. Thus, by ``soft'' we
mean soft compared to $m_b$, but still larger than $\Lambda_{QCD}$.
The jet
subprocess has typical momenta $p$  such that $p^{-} \gg  p^{+},p_{\perp}^{2}$
 with
$p^{+},p_{\perp}^{2} \sim O(m_{b}(1-x))$ and $p^{-} \sim O(m_b)$.
In order to delineate between momentum regimes, a factorization
scale $\mu$ is introduced. The fact that the process is $\mu$ independent
will be utilized to sum the large end point logs which are contained in the
soft and jet functions. The reduced digram for the inclusive radiative
decay $b\rightarrow X_s \gamma$ is exactly the same as above if we ignore the strange quark mass.

An important consequence of the factorization is the fact that 
the soft function, $S$, 
is universal. That is, it is independent of the final states as long as
factorization holds.
Thus the soft function in the semi-leptonic decay will be the
same soft function as in the radiative decay. This universality
will allow us to remove our ignorance of any nonperturbative
physics due to bound state dynamics by taking the appropriate ratio.
Thus, throughout this paper we will treat both the semi-leptonic as
well as the radiative decays in turn.

We conclude this discussion with a few comments. First,
we should point out  the differences between factorization in
the process considered here
and in deep inelastic
scattering for large values of the Bjorken scaling variable\cite{ct}. A crucial
difference arises from the fact that the initial quark is massive,
and hence, the semi-inclusive decays of the a heavy quark is infrared finite to
all orders in perturbation theory because there are no collinear divergences
arising from  initial state radiation. This fact has the important consequence
that the
differential decay rate will be independent of $\mu$. Whereas $\mu$ independence in deep inelastic scattering is only achieved after an appropriate
subtraction is made with another process, such as Drell-Yan, which
has the same collinear divergence structure as the deep inelastic scattering
process.
Next we note that, in general, the separation of diagrams into soft and
jet subprocesses is not unique, and some prescription must be adopted.
For a discussion of this issue see \cite{cs,sterman}. In our case, we will determine the proper separation from the
requirement of the $\mu$ independence of the decay rate from the condition 
that the hard scattering amplitude does not contain any large end point logs, and that
the purely collinear divergences in the jet must satisfy an Altarelli-Parisi
like  evolution equation. We will return to this point in the next section. 
The
factorization can be made more manifest by going to the light like axial gauge
with the gauge fixing vector pointing in the jet direction. In this gauge, the
soft lines decouple from the jet on a diagram by diagram basis.

In terms of the variables introduced earlier,
the triply differential factorized decay amplitude may be written as
\beq
{1\over{\Gamma^0}}{d^3\Gamma\over{dxdydy_0}}={6m_{b}(x-y)(y_0-x)}\int^{k^+_{max}}_{k 
^-_{min}}dk_+f(k^+,\mu^2)
J(p_h^-(p_h^+-k^+),\mu^2)H(m_b,\mu^2),
\eeq
\beq
\Gamma^0={G_F^2\over{96\pi^3}}\mid V_{ub}\mid^2m_{b}^5.
\eeq
This form will hold up to errors on the order of $O(1-x)$. 

We have chosen the electron to be traveling in the $+$ direction with momenta
$k=m_b(x,0,0_\perp)$, and
\beq
k^+_{max}=m_{b}(1-\eta),~~k^+_{min}=-(M_B-m_b).
\eeq
Here $f(k^+)$ is the probability for the $b$ quark to have light cone residual
momentum $k^+$, and thus contains not only  the information in the soft
function $S$ but also the non-perturbative information regarding the nature
of the bound state. If we ignore perturbative ``soft'' gluon radiation, then 
this function coincides with $\tilde{f}(k_+)$ defined in the previous section.
Notice that $k_+^{min}$ is negative. This is important non-perturbatively and represents
the leakedge past the partonic end point due to the soft gluon getting
energy and momentum from the light degrees of freedom inside the B meson. Loosely
speaking it is due to the Fermi motion of the b- quark inside the meson.  
 
Less formally, we may write the derivative of the decay amplitude as \cite{sk}

\begin{eqnarray}
{-1\over{\Gamma_0}}{d\over{dx}}\left({d\Gamma  
\over{dx}}\right)= 
\int^2_1dy_0 6(2-y_0)^2(y_0-1)G(x),
\end{eqnarray}
\beq
G(x)=\int_x^{M_B/m_b}dz
f(z,m_b/\mu)J(m_b^2(2-y_0)(z-x),\mu^2)H(m_b(2-y_0)/\mu).
\eeq 
In this equation we have changed variables from $k^+$ to the residual light
cone momentum fraction $z=(1-{k^+\over{m_b}})$, and absorbed a factor of
$m_b^2$ into the jet factor.

By taking the moments of this expression with respect to $x$ we see that we are able to treat  
the hard, soft and
jet functions separately. We are led to the following form for the
moments of the semi-leptonic rate
\begin{eqnarray}
\label{slmoms}
M_N^{sl}&\equiv & \frac{1}{\Gamma_0}\int^{M_B/m_b}_0 x^{N-1} 
 {d\over{dx}}\left({d\Gamma  
\over{dx}}\right)dx= \nonumber \\
& &\int^2_1dy_0 6(2-y_0)(y_0-1)
f_NJ_N(m_b^2(2-y_0),\mu^2)H(m_b(2-y_0),\mu),\\
&&f_N=\int^{M_B/m_b}_0z^Ndzf(z),\\
&&
J_N(2-y_0)=\int^1_0\lambda^NJ(m_b^2(2-y_0)(1-{\lambda},\mu^2))d\lambda.
\end{eqnarray}
In writing the last 
few equations we have dropped all terms of order $O(1-x)$ or  
equivalently taken
the large $N$ limit. The left hand side of Eq(\ref{slmoms}) defines the moments of the semi-leptonic decay electron distribution,
$M_{N}^{sl}$.

The moments of the soft function $f_N$ may be decomposed into a product
of moments of a perturbatively  
calculable $\sigma_N$ and the non-perturbative structure
function $S_N$, which corresponds to the moments of $\tilde{f}$ discussed in the
previous section. We may write 
\beq
\label{sft}
\tilde{f}(z)=\int^{M_B/m_b}_z{dy\over{y}}S(y)\sigma({z\over{y}}),
\eeq
and thus, taking moments,
\beq
f_N=\sigma_N S_N.
\eeq
An analogous situation exists for the decay $B \rightarrow X_{s} \gamma$.
We define 
$x= {2E_{\gamma}\over{m_b}}$
in the rest frame of the b-quark, and  take the photon to be moving in
the $+$  direction, and, as in the semi-leptonic decay,
at  $x \sim 1$ the hadronic jet is moving in the
$-$ direction. Furthermore, the invariant mass of the hadronic jet and its energy are
\beq
p_{h}^2= m_{b}^2(1-x), ~~~ p_{h}^0=m_{b}(1-x/2).
\eeq
Thus the s-quark is very energetic and since the invariant mass of the 
jet vanishes
as $x \sim 1$,  the s-quark  decays into quanta which are collinear
once we ignore effects on the order of $\frac{m_s^2}{m_b^2}$. Clearly the factorization picture
discussed earlier for the semi-leptonic decay holds here as well and the reduced
diagram is the same as in fig(3).

As before we may take the moments of the differential rate 
\beq
M_N^\gamma\equiv{1\over{\Gamma_{\gamma}}}\int^{M_B/m_b}_0 dx x^{N-1}{d\Gamma\over dx}= S_N \sigma_N 
J_N.
\eeq
where\cite{gsw},
\beq
\Gamma_{\gamma}= {{\alpha G_{F}^2}\over{32\pi^4}} m_{b}^5 \mid V_{tb}V_{ts}^{*} \mid^2
C_7^{2}(m_b).
\eeq
and
\beq
J_N= \int_0^1 dy y^{N-1}J(m_b^2(1-y),\mu^2).
\eeq
$\sigma_N$ and $S_N$ are the same functions defined in (\ref{sft}) and
$C_7$ is the Wilson coefficient of $O_7$ as defined in \cite{gsw}. 
For the radiative decay,
the $ln(1-x)/(1-x)_+$ distribution in the amplitude will correspond 
to $ln^2(N)$ in the moment. Whereas, in the semi-leptonic decay, taking
the derivative of the amplitude will generate plus distributions which
will then generate $LogN$ and $Log^2N$ after taking the moments.
Thus, we have reduced the problem of the resummation of the
large logs in the amplitude
 to resumming the logs in $J_N$ and $\sigma_N$ separately. This greatly 
simplifies the calculation as will be seen below.

\section{Resummation}  

The resummation of the infrared logs is analogous to summing
ultra violet logs. One takes advantage of the $\mu$ independence
of the amplitude. In the case of infra-red logs, $\mu$ is the factorization scale,
or equivalently the renormalization scale within the appropriate
effective field theory, which for this case would the  field theory
of Wilson lines \cite{kr}.

We first outline the derivation of a representation of the soft
function $\sigma_N$ near $x=1$ following the techniques developed in reference
 \cite{ct}. We will work in the
eikonal approximation where soft momenta are ignored wherever
possible.  At the one loop level, the real gluon emission
contribution factorizes and the quantity multiplying the tree level
rate is
\beq
F_{eik}^{real}(x)=g^2 C_F \int {d^3k\over (2\pi)^3 2k_0} \left({2p\cdot q\over p\cdot k
q\cdot k} - {m^2_b\over (p\cdot k)^2}\right)
\delta\left(1-x-{2k_0\over m}\right).  
\eeq
The $\delta$-function enforces the phase space constraint.
Similarly the one loop virtual gluon contribution is given by
\beq
F_{eik}^{virt}=-g^2 C_F \delta(1-x) \int {d^2k\over (2\pi)^3 2k_0} \left\{ {2p\cdot
q\over p\cdot k\> q\cdot k}- {m_b^2\over (p\cdot k)^2}\right\}.  
\eeq
Where $p$ and $q$  are the b quark and  light-quark momenta respectively.
In the Abelian theory exponentiation follows simply as a consequence
of the factorization in the eikonal approximation. For each gluon emission
one gets a factor of $F_{eik}^{virt}$ which is unitarized by 
the virtual contribution. After appropriate symmetrization the exponentiation
follows.
Next we use the result that even in a non-abelian theory, for the
semi-inclusive process under consideration, exponentiation of the one
loop result takes place \cite{gatheral,ft}.  By considering the $N^{\rm th}$
moment of the soft part, we obtain
\begin{eqnarray}
\sigma_N & = & \exp\Bigl\{g^2 C_F \int {d^3k\over (2\pi)^3 2k_0} 
\left[\left( 1-{2k_0\over M}\right)^{N-1} -1\right]\nonumber \\
& & \left[ {2p\cdot q\over p\cdot k\> q\cdot k} - { m_b^2\over (p\cdot
k)^2} \right]\Bigr\}.
\end{eqnarray}

It should be noted that the ultraviolet cutoff is determined by the factorization scale $\mu.$ This cutoff is necessary despite the fact that the process
under consideration is infrared finite. All momentum above this scale
get shuffled into the  hard scattering amplitude $H$.
The need for a cut off 
stems from the
fact that we have used the eikonal approximation. This approximation
is equivalent to a Wilson line formulation of the problem, and thus, 
as in heavy quark effective field theory, generates a new velocity
dependent anomalous dimension \cite{kr}.

By an appropriate change of variables $\sigma_N$ may be written as
\begin{eqnarray}
\sigma_N &=& \exp\Bigl\{-\int^{\mu/m_b}_0 {dy\over y} 
\left(1-(1-y)^{N-1}\right)\nonumber \\
& & (\int^{\mu^2}_{m_b^2y^2} {dk_\perp^2\over k_\perp^2} C_F
{\alpha_s\over \pi} (k_\perp^2) - C_F {\alpha_s\over \pi}
(m_b^2y^2))\Bigr\}.
\end{eqnarray}
In arriving at the above, we have made the replacement $\alpha_s
\rightarrow \alpha_s (k_\perp^2)$. This  change  has
the effect of resumming the next to leading logs coming from
collinear emission of light fermion pairs\cite{amati}. However, it does
not sum all the soft sub-leading logs.

Explicit calculations carried out at the two loop level \cite{kod,ct}
indicate that the  rest of the sub-leading terms in the above may be included \cite{ct}
\beq
{C_F\alpha_s (k^2_\perp)\over \pi}\longrightarrow A (\alpha_s (k^2_\perp)),
\eeq
with
\begin{eqnarray}
A(\alpha_s) &= & {\alpha_s\over \pi} C_F + \left({\alpha_s\over
\pi}\right)^2 {1\over 2} C_F k,\nonumber \\ k &=&
C_A\left({67\over 18} - {\pi^2\over 6}\right) - {10\over 9} T_R N_f.
\end{eqnarray}
This resums all the leading and next to leading logs of $N$ in the
soft function.

Thus, we obtain  
\begin{eqnarray} \sigma_N (m_b/\mu) &=&
\exp\Bigl\{-\int^{\mu/m_b}_0 {dy\over y} (1-(1-y)^{N-1})\nonumber\\
& &(\int^{\mu^2}_{m_b^2y^2} {dk_\perp^2\over k_\perp^2} A\left(\alpha_s
(k_\perp^2)\right) + B\left(\alpha_s(m_b^2y^2\right))\Bigr\},
\end{eqnarray}
with $B(\alpha_s) = -\alpha_s/\pi.$
This integral is not well defined due to the existence of the Landau pole, 
and a prescription is needed to define the integral. Choosing a prescription
leaves an ambiguity on the order of the power corrections \cite{sk}\cite
{ben95}\cite{akh95}.
If 
we use  the large $N$ identity
\begin{eqnarray}
\label{largeN}
1-x^{N-1}&=& \theta\left(1-x-{1\over \tilde{N}}\right), \nonumber \\
\tilde{N}&=&\frac{N}{N_0}, ~~~~N_0=e^{-\gamma_E},
\end{eqnarray}
which is accurate to within $2\%$ at $N=10,$
to rewrite 
\begin{eqnarray} \sigma_N (m_b/\mu) & = & \exp
\Bigl\{-\int^\mu_{m_b/\tilde{N}} {dk_\perp\over k_\perp} \Bigl[2A\left(\alpha_s
(k_\perp)\right) \ln {k_\perp \tilde{N}\over m_b}
 -B \left(\alpha_s (k_\perp^2)\right)\Bigr],
\end{eqnarray}
then we have fixed a prescription which is unambiguous to the accuracy
we are concerned with in this paper.
From this result, we find that $\sigma_N(m_b/\mu)$ satisfies the RG
equation
\vskip.1in
\begin{eqnarray}
\label{rg}
\left(\mu {\partial\over \partial\mu} + \beta {\partial\over
\partial g}\right)\sigma_N \left({m_b\over\mu}\right) 
= - \left[2A \left(\alpha_s(\mu^2)\right) \ln {\mu \tilde{N}\over m_b} + B
\left(\alpha_s (\mu^2)\right)\right]\sigma_N \left({m_b\over\mu}\right).
\end{eqnarray}
We note that this is in agreement with reference\cite{sk} where
Wilson line techniques were utilized.

We may now use the $\mu$ dependence of the soft function, together with
the fact that the total amplitude is $\mu$ independent, to
determine the renormalization group equation satisfied by the
jet and hard functions.
We have seen in section (3) that the $N^{th}$ of the  derivative of
the moments of the semi-leptonic decay  has the factorized form
\beq
\sigma_N (m_b\sqrt{2-y_0}/\mu) J_N^{sl} (m_b/\mu) H^{sl} (m^2_b(2-y_0)/\mu^2).
\eeq
 Whereas, for the radiative decay the moments of the decay spectrum is given by
\beq
\sigma_N (m_b/\mu) J_N^\gamma (m_b/\mu) H^\gamma (m^2_b/\mu^2).
\eeq
We have now labeled the jet and hard functions according to their
processes since these function are not universal.
 We will first consider the RG
equation satisfied by  $J_N^{\gamma}$ and $H^\gamma$.
The equations satisfied by $J^{sl}$ and $H^{sl}$ can then
be determined  by simply by making  the appropriate replacements.
We may derive the RG  equations satisfied by
these functions by using the following facts
\begin{itemize}
\item $\mu {d\over d\mu} (\sigma_N (m_b/\mu) J_N (m_b/\mu) H(m_b/\mu)) = 0$
\item The RG equation satisfied by $\sigma_N (m_b/\mu)$ is given by Eq. (\ref{rg})
\item The hard scattering amplitude by definition has no $N$ dependence 
\item The jet functional form which is
$
J_N \left( {m_b^2\over \tilde{N}} \cdot {1\over \mu^2}\right),
$
\end{itemize}
Leading to the following RG equations
\begin{eqnarray}
\left(\mu{\partial\over \partial\mu} + \beta{\partial\over \partial g} 
+f (\alpha_s)\right)J_N \left(m_b/\mu\right)
= 2A (\alpha_s (\mu^2))\ln {\mu^2\tilde{N}\over m_b^2} J_N (m_b/\mu),
\end{eqnarray}
\begin{eqnarray}
\left(\mu{\partial\over \partial\mu} + \beta{\partial\over \partial g} -
f (\alpha_s) - B(\alpha_s (\mu^2))\right)H(m_b/\mu)
= -2A \left(\alpha_s (\mu^2)\ln {\mu\over m}\right) H (m_b/\mu).
\end{eqnarray}
$f(\alpha_s)$ is an arbitrary function which can only be determined
from additional input. We fix $f(\alpha_s$) by requiring that the purely collinear divergences
of the jet factor be determined by an Altarelli-Parisi type equation
as discussed in \cite{ct,sterman}.  We note that for these purposes the jet
factor is a cut light quark propagator in the axial gauge.

By requiring that we correctly reproduce
the pure collinear divergences
at the one-loop level it is found that
\beq
f(\alpha_s) = \ 2\gamma (\alpha_s).
\eeq
Where $\gamma (\alpha_s)$ is the axial gauge anomalous dimension
\cite{ct,sterman} 
\beq
\gamma(\alpha_s) = - {3\over 4} {\alpha_s\over \pi} C_F + \ldots .
\eeq
The solution of the jet RG equation may be written 
(to the desired accuracy) in the form
\begin{equation}
J_N^{\gamma} (m_b/\mu) = \exp \left\{
\int^{\mu^2/m_b^2}_{1\over \tilde{N}} {dy\over y}\left[\int^{\mu^2}_{m_b^2y} 
{dk_\perp^2\over k_\perp^2} A(\alpha_s (k_\perp^2)) - \gamma (\alpha_s 
(m_b^2 y))\right]\right\}.
\end{equation}
For future purposes, we rewrite this
in the form
\begin{eqnarray}
J_N^{\gamma} (m_b/\mu) &=&\exp\left\{ \int^1_0 {dy\over y} 
\left[1-(1-y)^{N-1}\right] \left[\int^{\mu^2}_{m_b^2y}
{dk_\perp^2\over k_\perp^2}
A(\alpha_s(k_\perp^2))-\gamma(\alpha_s(m_b^2 y))\right] \right.\nonumber\\
&& \left.+ \int_1^{\mu^2/m_b^2} {dy\over y} \left[\int^{\mu^2}_{m_b^2y}
{dy^2\over k_\perp^2} A(\alpha_s(k_\perp^2)) - \gamma(\alpha_s(m_b^2 y))\right]
\right\}.\end{eqnarray}

We may now write the explicit expressions for the resummed jet and soft
factors. For  the radiative decay $B\rightarrow X_s \gamma$
we rewrite the various representations obtained earlier leaving
\begin{equation}
\sigma_N (m_b/\mu) = \exp \int^1_0 {dz\over 1-z} (1-z^{N-1})N(z),
\end{equation}
\vskip.05in
\begin{equation} 
J_N (m_b/\mu) = \exp
 \int^1_0 {dz\over 1-z} (1-z^{N-1})I(z),
\end{equation}
\vskip.05in
\begin{eqnarray}
N(z)&=&\int^{\mu^2}_{m_b^2(1-z)^2} \Bigl[{dk_\perp\over k_\perp^2} -A(\alpha_s
(k_\perp^2))\Bigr]- B (\alpha_s (m_b^2 (1-z)^2)) 
\nonumber \\
&&\qquad -\int_1^{\mu/m_b}
\left[\int^{\mu^2}_{m_b^2y^2} {dk_\perp^2\over k^2_\perp} A(\alpha_s
(k_\perp^2)) + B(\alpha_s(m_b^2 y^2))\right],
\end{eqnarray}
\vskip.05in
\begin{eqnarray}
I(z)&=&\Bigl[
\int^{\mu^2}_{m_b^2(1-z)} {dk_\perp^2\over k_\perp^2}
A(\alpha_s(k_\perp^2))\Bigr] -\gamma (\alpha_s (m_b^2 (1-z))) 
\nonumber \\
&&\qquad +\int^{\mu^2/m_b^2}_1 {dy\over y} \left[ \int^{\mu^2}_{m_b^2y}
{dk_\perp^2\over k_\perp^2} A(\alpha_s (k_\perp^2))- \gamma(\alpha_s (m_b^2y))
\right].
\end{eqnarray}

Combining these two factors we see that for the $N$ dependent piece in
the exponent, the $\mu^2$ dependence exactly cancels.  There are,
however, pieces which are independent of $N$ which are $\mu$ dependent and these will
combine with similar terms in the hard scattering amplitude to give a
$\mu$-independent answer which must be true by construction.
Combining all the factors we find
\begin{equation}
\sigma_N (m_b/\mu) J_N (m_b/\mu)=
\exp -\int^1_0 {dz\over 1-z}
(1-z^{N-1})K(z),
\end{equation}
\vskip.05in
\begin{equation}
K(z)=\int^{m_b^2(1-z)}_{m_b^2(1-z)^2} {dk_\perp^2\over
k_\perp^2} A(\alpha_s (k_\perp^2))
- \gamma (\alpha_s(m_b^2 (1-z))) - B (\alpha_s (m_b^2 (1-z)^2))\Bigr].
\end{equation}
The $N^{th}$ moment of the decay rate in then given by
\beq
M_{N}^{\gamma}= S_{N}\sigma_{N} J_{N}^\gamma H^\gamma(\alpha_s(m_b^2)).
\eeq
The value of the the one loop 
hard scattering amplitude $H^\gamma$  is given in the Appendix.

For the case of the semi-leptonic decay 
the expression for the soft factor is the same as above. However,
for the jet, we must rescale
$J_N^{sl} ( m_b/\mu) \rightarrow J_N \left(m_b/\mu \sqrt{2-y_0}\right).$  
Thus, we get
\begin{equation}
J_N^{sl} \left( m_b/\mu \sqrt{1-x_\nu}\right)=
exp \Biggl\{ \int^1_0 {dy\over y} \left[ 1-(1-y)^{N-1}\right] 
L(y,x_\nu)\Biggr\},
\end{equation}
\begin{equation}
 L(y,x_\nu)=\int^{\mu^2}_{m_b^2y(1-x_\nu)}
{dk_\perp^2\over k_\perp^2}
A(\alpha_s(k_\perp^2))-\gamma(\alpha_s(m_b^2y(1-x_\nu))).
\end{equation}
In writing the above, we have used the fact that $y_0=x+x_\nu$, and $x \sim 1$
to replace the variable $y_0$ by $x_\nu$, the neutrino energy fraction.
After some algebra the above may be combined with the expression for the
perturbative soft function, such that for the product we may write

\begin{equation}
\label{sl}
\sigma_N (\frac{m_b}{\mu}) J_N^{sl} (\frac{m_b}{\mu} \sqrt{1-x_\nu})
=\exp \Biggl\{ - \int^1_0 {dz\over 1-z} (1-z^{N-1})Q(z)+P(N,x_\nu)\Biggr\},
\end{equation}
\begin{equation}
Q(z)=\int^{m_b^2(1-z)}_{m_b^2(1-z)^2} {dk_\perp^2\over k_\perp^2}
A(\alpha_s (k_\perp^2)) 
-\gamma (\alpha_s( m_b^2 (1-z)))
- B(\alpha_s (m_b^2(1-z)^2)),
\end{equation}
\begin{equation} 
P(N,x_\nu)= \int^1_{1/\tilde{N}} {dy\over y} \int^{m_b^2y}_{m_b^2y(1-x_{\nu})}
{dk_\perp^2\over k_\perp^2} A(\alpha_s (k_\perp^2))
- \int^1_{1-x_\nu} {dy\over y} \left[\gamma \left(\alpha_s
\left({m_b^2y\over \tilde{N}}\right)\right) - \gamma (\alpha_s (m_b^2y))\right].
\end{equation}

We note that in deriving this form we have kept only the $N$ dependent
pieces in the exponent.  Our analysis shows that certain $N$ independent
terms, like those proportional to $\ln (1-x_\nu)$, can also be resummed
using the above mentioned procedure.  However, we have taken $\alpha_s
(m_b^2)\ln (1-x_\nu)$ to be small in the relevant $x_\nu$ range and
hence relegated all of these logs to the
hard scattering amplitude.
Thus, we may write for the $N^{th}$ moment of the 
semi-leptonic decay
rate, up to corrections $O(1/N)$, as
\beq
\label{64}
M_{N}^{sl}=
S_{N}\int^1_0dx_{\nu} 6 x_{\nu}(1-x_{\nu})
\sigma_NJ_N(m_b^2(1-x_\nu),\mu^2)H(m_b(1-x_\nu),\mu).
\eeq
The complete expression for the one loop 
hard scattering amplitude both the  radiative and semi-leptonic processes 
at $x \sim 1$ are given in the Appendix.

We conclude this section by giving some simplified expressions for the product
$\sigma_N J_N$ which will be useful for numerical analysis.  
We begin by noticing that,  as long as 
$\alpha_s(m_b^2)lnN \leq 1$, the resummation formulae given above have
a convergent power series expansion in $\alpha_s(m_b^2)lnN$ to the next to 
leading log accuracy. Thus, we compute these expressions to this accuracy
and delegate all the non perturbative effects phenomenologically to the
structure function $S_N$. For a similar approach for the case of
$e^+e^-$ annihilation see \cite{web}.
To evaluate the integrals in the exponent, we may perform the $z$
integration using the large $N$ identity (\ref{largeN}).   
The $k_{\perp}$ integration is simplified by using the RG equation for the running coupling to change variables to $\alpha_s$, i.e.
\beq
{dk_{\perp}^2 \over k_{\perp}^2} = -{1\over \beta_0}{d\alpha_s \over \alpha_s^2}
(1-{\beta_1 \over \beta_0}\alpha_s+O(\alpha_s^2))
\eeq
where
\beq
\beta_0={{11C_A-2N_f} \over 12\pi}, ~~~ \beta_1={{17C_A^2-5C_AN_f-3C_FN_f} \over 24\pi^2}.
\eeq
Next we use the expansion, correct to next to leading log accuracy,
\beq
\alpha_s(m_b^2/N)={\alpha_s(m_b^2) \over {1-\beta_0\alpha_s(m_b^2)lnN}}
\left(1-{\beta_1 \over \beta_0}{\alpha_s(m_b^2) \over {1-\beta_0\alpha_s(m_b^2)lnN}}
ln(1-\beta_0\alpha_s(m_b^2)lnN)\right),
\eeq
to obtain 
\beq
\label{result}
ln(\sigma_NJ_N)=lnN\left(g_1^\gamma(\chi)\right)+g_2^\gamma(\chi),
\eeq
where,
\beq
\chi= \beta_0\alpha_s(m_b^2)lnN.
\eeq
In the above, the functions $g_1$ and $g_2$ have the following form for the two
processes discussed in this paper

For the radiative decay
\beq
g_1^{\gamma}=-{A^{(1)} \over{2\pi\beta_0\chi}}((1-2\chi)ln(1-2\chi)-2(1-\chi)ln(1-\chi)),
\eeq
and
\begin{eqnarray}
g_2^{\gamma}&=&-{A^{(2)} 
\over{2\pi^2\beta_0^2}}\left(-ln(1-2\chi)+2ln(1-\chi)\right)
-{{A^{(1)}\beta_1} 
\over{2\pi\beta_0^3}}\Bigl( ln(1-2\chi)
-2ln(1-\chi)  \nonumber \\
&+&  \frac{1}{2} ln^2(1-2\chi)-ln^2(1-\chi)\Bigr)
+{\gamma^{(1)} 
\over{\pi\beta_0}}ln(1-\chi) +{B^{(1)} \over{2\pi\beta_0}}ln(1-2\chi)\nonumber
\\
&-&\frac{A^{(1)}}{\pi \beta_0}lnN_0(ln(1-2\chi)-ln(1-\chi)).
\end{eqnarray}

For the semi-leptonic decay
\beq
g_1^{sl}= g_1^{\gamma},
\eeq
and
\beq
g_2^{sl}= g_2^{\gamma}+ {A^{(1)} \over{\pi\beta_0}}ln(1-x_{\nu})ln(1-\chi).
\eeq
 We have kept only the $N$ dependent terms in these $g$ factors which
 exponentiate. There are
also $N_0$ dependent constant terms which we will shuffle into the
hard scattering amplitude. These terms are given by
\begin{eqnarray}
h^\gamma&=&\frac{\alpha_s}{\pi}lnN_0(B^{(1)}+g^{(1)})-\frac{\alpha_s}{2\pi}
ln^2N_0, \nonumber \\
h^{sl}&=&h^\gamma+A^{(1)}\frac{\alpha_s}{\pi}lnN_0ln(1-x_\nu).
\end{eqnarray}
Furthermore eq. (\ref{64}) for the semi-leptonic case becomes
\beq
\label{slrate}
-\frac{1}{\Gamma_0}\int_0^1x^{N-1}\frac{d}{dx}\frac{d\Gamma}{dx}=
S_N\int_0^1dx_\nu 6(1-x_\nu)x_{\nu}(H(x_\nu)+h^{sl})Exp(g_1+g_2^{sl}).
\eeq
In writing the above, we have used the notation
\beq
A(\alpha_s)=({\alpha_s \over \pi})A^{(1)}+({\alpha_s \over \pi})^2A^{(2)},
\eeq
and
\beq
B(\alpha_s)=({\alpha_s \over \pi})B^{(1)}, ~~\gamma(\alpha_s)=
({\alpha_s \over \pi})\gamma^{(1)}.
\eeq
The values for $A^{(1)},A^{(2)},B^{(1)},\gamma^{(1)}$ have been given previously. 
$H(x_\nu)$ is the one loop correction to the hard scattering amplitude
given in the Appendix.

It is interesting to note that if we expand the expressions for 
$g_1^{sl}$ and $g_2^{sl}$ in (\ref{result}), we see that $G_{24}$, as defined
in (\ref{exp}), 
vanishes. Thus, the two loop results does not trivially exponentiate
as one might have naively thought. Such behavior is a universal
property of the asymptotic limit of distribution functions and is a consequence of the fact that only ``maximally non-abelian'' graphs contribute to the
exponent beyond one loop\cite{gatheral}. Knowing this greatly reduces
the number of graphs that need to be calculated in a general resummation
procedure.

From expression (\ref{slrate}) we may determine the range of $N$ for which
our calculation is valid.
The integration over y contains a branch cut at
$N=\frac{m_b}{\Lambda_{QCD}}$, signaling the breakdown of the
the perturbative formalism. This breakdown is coming from the
fact that the time scale for gluon emission is becoming too long.
An inspection of the resummation formulae for these quantities suggests
that in the region $z \sim 1$ such that $k_{\perp}^2 \leq \Lambda^2$ ,
non-perturbative effects become important. Thus, we conclude that we
may only trust our results in the range
\beq
N<\frac{m_b}{\Lambda_{QCD}}.
\eeq

\section{Analysis and Results}

\begin{figure}
\centering
\epsfysize=2.5in   
\hspace*{0in}
\epsffile{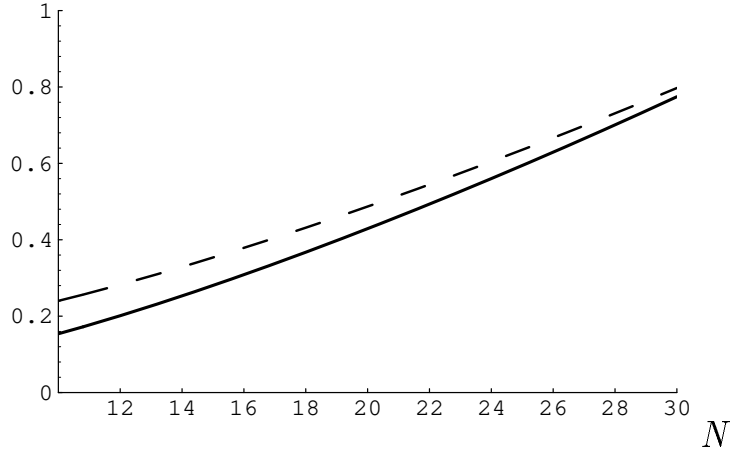}  
\caption{The difference between the moments of the one loop
result and the resummed result with (dashed) and without (undashed)
the resummation of the $\pi^2$, normalized to the one loop result.
$N$ varies from $10$ to $30$.}
\label{rsult}  
\end{figure}

With the resummation now in hand, let us consider the relative sizes of all
the contributions. In figure \ref{rsult} we show the difference between the
one loop result and the resummed rate given by eq.(\ref{slrate}) normalized to
the moments of the one loop result,
\beq
\label{difone}
-\frac{1}{\Gamma_0}\int_0^1x^{N-1}\frac{d}{dx}\frac{d\Gamma}{dx}=
1-\frac{2\alpha_s}{3 \pi}(\pi^2+\frac{5}{4}-\frac{31}{6}ln\tilde{N}+ln^2\tilde
{N}).
\eeq
In our calculation we take  $\Lambda_{QCD}^{n_f=4}\approx 200~MeV$.
We see that resumming the next to leading logs 
has a $20\%-50\%$ effect in the range of $N$ we are considering.
Furthermore, for completeness we have included the effect of
resumming the $\pi^2$. The result of this resummation is given
by the dashed line. We see that the effect of resumming the 
$\pi^2$ is small\footnote{Note that
in resumming the $\pi^2$ we only resum part of the $\pi^2$ in the
expression (2), since part of the $\pi^2$ contribution comes from
integration over the neutrino energy}. 
As a check of the numerics we compared the
resummed expression to the one loop result (\ref{difone})
and found that for small $N$ the two coincide to within less than a percent. 
The fact that the resummation of the next to leading logs is more
important that the leading logs, is rather disheartening.
It leads one to believe that perhaps the next to next to leading
logs will be even more important. However, the fact that the effect of 
subleading logs is larger than the leading is already hinted at one
loop, given that the ratio of the coefficients in front of these
logs is $31/6$. It could be {\it hoped}  that the ratio of the coefficients
of the next to leading and next to next to leading logs is not
so large and the terms left over in our resummation will be
on the order of $10\%$.

\section{Discussion}

Before we conclude with a discussion of the future prospects of the extraction
of $V_{ub}$ we wish to point out that there is one tacit assumption 
which has been made up to this point in our investigation. That is, we have assumed that 
local duality will hold when we are a few hundred MeV from the end point.
The whole formalism of using the OPE in calculating
inclusive decay rates assumes that at certain parts of the Dalitz plot,
the Minkowski space calculation will give the correct result.
This should be a good approximation as long as we stay away from
the resonance region. The question is, how far from the end point does
this region begin? If it is found that single resonances dominate, 
even as far as a few hundred MeV from the endpoint, then the extraction
of $V_{ub}$ through inclusive decays is surely doomed. The quark model
seems to indicate that this may be the case \cite{isgw}, 
though other theoretical
predictions say otherwise.  
We will have to wait to see the data before we can decide on the fate
of the extraction methods discussed here.

Next we wish to reiterate that ${\it completely}$ eliminating the background
from $b\rightarrow c$ transition by going to very large $N>30$,
is not feasible since there is no way to reliably calculate the
soft gluon emission which takes place. This is because when one
goes that far out on the tail, the time scale for gluon emission is
too long compared to the QCD scale to have any hope of perturbation theory making any sense.
Again, this statement is independent of how many soft logs one is
willing to resum. Another way of saying this is that when $x$ gets to
close to one, there is no operator product expansion since the
expansion parameters is $\frac{\Lambda}{m_b(1-x)}$.
Thus, we are stuck with the fact that there will
always be contamination from transitions to charmed final states.
Calculating the end point of the charmed spectrum using the techniques
discussed above fail as well since resonances will dominate.
Thus, there does not seem to be any way to avoid having to
use a model to determine the background in the extraction process.
The best we can hope to
do, using the results in this paper, is to go to a large 
enough  value of $N$ that we can reduce the model dependence
as much as possible. Certainly, we can greatly reduce the model dependence
from what it is in present extractions which rely solely on models.

The last point that needs to mentioned is the fact that measuring
large moments it not experimentally feasible, as $x^N$ varies
much too rapidly. For instance, if we assume that the bin size
is given by $\delta=\frac{\delta E}{m_b}$, then  the error at point
$x$ for the $N$th moment will be $\frac{\delta N}{x}$. Therefore, the
error can accumulate quite rapidly. Thus, it will be necessary to take
the Mellin transform of our result. Given that our result is only
trustable for $N<25$, one must be careful to calculate the contribution
to the inverse transform from higher moments, if one hopes to impose
the bounds on the errors discussed in this paper. Also, for smaller
values of $N$ one must be sure not to use the resummed formula as
we have dropped terms that go like $1-x$.
 
Given these caveats, we  may now address the issue as to what accuracy we
can determine $V_{ub}$ using inclusive decays. 
Since the sub-leading logs dominate the leading logs, the conservative 
conclusion would be that a model independent extraction of $V_{ub}$
is not possible. However, let us proceed under the assumption the sub-sub-leading logs will be smaller than the sub-leading logs. In this 
 case we may say that we have been able to reduce
the errors from radiative corrections down to the order of $10\%$.
However, the QCD perturbative expansion is notoriously asymptotic, and though
we may hope that we have resummed the dominant pieces of the expansion,
there could still be large constants (independent of $N$) which could arise. 
 
Another source of errors will come from the fact that we need to eliminate
the dependence of the decay rate on the moments of the non-perturbative
structure function \cite{neub} by taking the ratio of the semi-leptonic
decay moments with the moments of the radiative decay. This will introduce
the errors in the radiative decay into the semi-leptonic decay. 
One could calculate without any non-perturbative resummation, thus eliminating
these errors (the results in this paper are easily modified to include this
possibility), but then it is difficult to quantify the model dependent errors
introduced in the truncation. 
Finally, there are the errors introduced due to the model dependence
from the calculation of  the background. This error will be reduced
as we choose larger values of $N$. This is the most difficult error to
quantify, and we shall not discuss it here. 

The authors believe that,
if the end point is not dominated by single particle resonances, and if we
assume that the fact that the sub-leading logs dominate the leading
logs is just an anomaly,  then we
may hope to eventually extract $V_{ub}$ at the $30\%$ level using the 
results presented here. 
Moreover, resumming the next to leading logs is indeed necessary.
However, the more conservative view would be that
the  endpoint calculation is just intractable at this time
, since it could be that
the sub-sub-leading logs will dominate. To be sure that this is not
the case the sub-sub-leading logs would need to be resummed. This would
entail calculating $A$ to three loops, and $B$ and $\gamma$ to  two loops.
Without this calculation, we can not determine with certainty the size of the
errors.

\section {Appendix}

In this Appendix we give explicit expressions for the hard scattering amplitude
at the one loop level and to leading order in $(1-x)$.
We first present the results of the computation of the QCD corrections to the
doubly differential rate ${d^{2}\Gamma} \over {dxdy_0}$ for the semi-leptonic
b-quark decay. It is clear from sections 2,3 that this is the 
quantity whose moments factorize, and which is relevant for the resummation. 
We write,
\beq
{{d^{2}\Gamma} \over {dxdy_0}}= 6{\Gamma_0}(y_0-1)(2-y_0)(1- 
{{2\alpha_s}\over{3\pi}}G(x,y_0)).
\eeq
where, $\Gamma_0$  was defined earlier.
The contributions of
the real and the virtual gluon emission diagrams  to $G(x,y_0)$ are given by
\begin{eqnarray}
G^{real}_{fin}&=&ln^{2}(1-x) + \frac{7}{2}ln(1-x)
-2ln(2-y_0)ln(1-x)-ln^{2}(2-y_0)+\frac{3}{2}ln(2-y_0)+\frac{\pi^2}{6}
\nonumber \\
&-&\frac{1}{2}ln^2({{\lambda^2}\over m_{b}^2})
-\frac{5}{2}ln({{\lambda^2}\over m_{b}^2})
+2ln(2-y_0)ln({{\lambda^2}\over m_b^2}),
\end{eqnarray}
and
\begin{eqnarray}
G^{virt}_{fin}&=& 
 3ln^2(2-y_0) - 2ln(2-y_0)ln(y_0-1)
+2Re(Li_2({1\over{2-y_0}}))
+3{(2-y_0)\over{y_0-1}}ln(2-y_0)\nonumber
 \\ &-&2{ln(2-y_0)\over{y_0-1}}
+\frac{5}{2}+\frac{\pi^2}{6}+\frac{1}{2}ln^2({\lambda^2\over {m_b^2}})
+ \frac{5}{2}ln({\lambda^2\over {m_b^2}})
\nonumber \\
&-&2ln(2-y_0)ln({\lambda^2\over {m_b^2}}).
\end{eqnarray}
In the above, $\lambda$ is the gluon mass used to regulate the infrared
divergences at intermediate stages of the calculation.
Combining these results and integrating over $y_0$ gives the electron spectrum
which agrees with \cite{jk} but disagrees with \cite{corbo}.

From this we see that to the approximation we are working in, the hard scattering
amplitude as defined in eq.\ref{slmoms} is given by
\begin{eqnarray}
H_{sl}&=& 1- {{2\alpha_s}\over{3\pi}}\left(\frac{\pi^2}{3} + 2ln^{2}(2-y_0) - 2ln(2-y_0)ln(y_0-1)
-\frac{3}{2}ln(2-y_0) \right. 
\nonumber \\
 &+& \left. {{ln(2-y_0)}\over{y_0-1}}+2Re(Li_2({1\over{2-y_0}}))
+\frac{5}{2} \right).
\end{eqnarray}

For the radiative decay, we may extract the hard scattering amplitude from
\cite{aligreub}
\beq
H_{\gamma}= 1-{{2\alpha_{s}}\over {3\pi}}(\frac{3}{2}- \frac{2\pi^2} {3}).
\eeq

\vskip.5in

\centerline{\bf {Acknowledgments}}
The authors would befitted from discussions with: G. Boyd, G. Korchemsky, 
Z. Ligeti,
0 M. Luke, A. Manohar, L. Trentadue,  and especially M. Beneke, A. Falk, G. Sterman 
 and M. Wise. R.A. Would like to thank C.E.A. Saclay, where part of this work was done, for their hospitality.

\end{document}